\documentclass[12pt]{article}
\usepackage{times}
\usepackage{geometry}
\usepackage[utf8]{inputenc}
\usepackage{enumitem,amssymb}
\usepackage{ragged2e}
\usepackage{graphicx}
\usepackage{sidecap}
\usepackage{natbib}
\usepackage{aas_macros}
\newlist{thematic}{itemize}{8}
\setlist[thematic]{label=$\square$}
\usepackage{pifont}
\newcommand{\cmark}{\ding{51}}%
\newcommand{\done}{\rlap{$\square$}{\raisebox{2pt}{\large\hspace{1pt}\cmark}}%
\hspace{-2.5pt}}

\let\ga=\gtrsim
\newcommand\arcmin{\mbox{$^\prime$}}%
\newcommand\arcsec{\mbox{$^{\prime\prime}$}}%

\newcommand{\kms}         {km~s$^{-1}$}

\newcommand{\msun}        {M$_{\odot}$}

\newcommand{\lcdm}        {$\Lambda$CDM}

\begin{document}
\pagestyle{empty}
\raggedright
\huge
Astro2020 Science White Paper \linebreak

Testing the Nature of Dark Matter with Extremely Large Telescopes \linebreak
\normalsize

\noindent \textbf{Thematic Areas:} \hspace*{60pt} $\square$ Planetary Systems \hspace*{10pt} $\square$ Star and Planet Formation \hspace*{20pt}\linebreak
$\square$ Formation and Evolution of Compact Objects \hspace*{31pt} $\done$ Cosmology and Fundamental Physics \linebreak
  $\square$  Stars and Stellar Evolution \hspace*{1pt} $\square$ Resolved Stellar Populations and their Environments \hspace*{40pt} \linebreak
  $\square$    Galaxy Evolution   \hspace*{45pt} $\square$             Multi-Messenger Astronomy and Astrophysics \hspace*{65pt} \linebreak
  
\textbf{Principal Author:}

Name: Joshua D. Simon
 \linebreak						
Institution: Carnegie Observatories
 \linebreak
Email: jsimon@carnegiescience.edu
 \linebreak

\justify

\textbf{Co-authors:} Simon Birrer (UCLA), Keith Bechtol (University of Wisconsin-Madison), Sukanya Chakrabarti (RIT), Francis-Yan Cyr-Racine (Harvard University \& University of New Mexico), Ian Dell'Antonio (Brown), Alex Drlica-Wagner (Fermilab), Chris Fassnacht (UC Davis), 
Marla Geha (Yale), Daniel Gilman (UCLA), Yashar D. Hezaveh (Flatiron Institute), Dongwon Kim (UC Berkeley), Ting S. Li (Fermilab), Louis Strigari (Texas A\&M), and Tommaso Treu (UCLA)\linebreak

\noindent
\textbf{Abstract:} For nearly 40 years, dark matter has been widely
assumed to be cold and collisionless.  Cold dark matter models make
fundamental predictions for the behavior of dark matter on small
($\lesssim10$~kpc) scales.  These predictions include cuspy ($\rho
\propto r^{-1}$) density profiles at the centers of dark matter halos
and a halo mass function that increases as $dN/dM \propto M^{-1.9}$
down to very small masses.  We suggest two observational programs
relying on extremely large telescopes to critically test these
predictions, and thus shed new light on the nature of dark matter.
(1) Combining adaptive optics-enabled imaging with deep spectroscopy
to measure the three-dimensional motions of stars within a sample of
Local Group dwarf galaxies that are the cleanest dark matter
laboratories known in the nearby universe.  From these observations
the inner slope of the dark matter density profile can be determined
with an accuracy of 0.20~dex, enabling a central cusp to be
distinguished from a core at $5\sigma$ significance.  (2)
Diffraction-limited AO imaging and integral field spectroscopy of
gravitationally lensed galaxies and quasars to quantify the abundance
of dark substructures in the halos of the lens galaxies and along the
line of sight.  Observations of 50 lensed arcs and 50 multiply-imaged
quasars will be sufficient to measure the halo mass function over the
range $10^{7} < M < 10^{10}$~M$_{\odot}$ at cosmological scales,
independent of the baryonic and stellar composition of those
structures.  These two observational probes provide complementary
information about the small scale structure, with a joint
self-consistent analysis mitigating limitations of either probe.  This
program will produce the strongest existing constraints on the
properties of dark matter on small scales, allowing conclusive tests
of alternative warm, fuzzy, and self-interacting dark matter models.

\pagebreak
\pagestyle{plain}
\setcounter{page}{1}

\section{The Astrophysics of Dark Matter}
\vspace{-0.4cm}

The nature of dark matter remains one of the most important questions
in physics. Because of the tremendous overall successes of the dark
energy + cold dark matter ($\Lambda$CDM) model, dark matter is
widely assumed to consist of a massive, weakly-interacting
particle. However, decades of effort to detect such a particle with
accelerators, direct detection experiments, and indirect searches for
the products of dark matter annihilation or decay have thus far failed
to identify any signature of the dark matter particle. At present,
astrophysical measurements still represent our only means of directly
studying the properties of dark matter.

\vspace{-0.5cm}
\subsection{Problems on Small Scales}
\vspace{-0.2cm}

Although the large-scale structure of the universe is in excellent
agreement with the predictions of the $\Lambda$CDM paradigm, numerical
simulations have identified possible discrepancies with the observed
properties of galaxies on small scales ($<10$~kpc). In particular, in the absence of self-interactions or
interactions with baryons, halos composed of massive dark matter particles should
follow a universal density profile, with $\rho \propto r^{-1}$ at
small radii and $\rho \propto r^{-3}$ at large radii~\citep[e.g.,][]{Navarro1996,Navarro2004}. However, observations of dark matter-dominated
galaxies and galaxy clusters generally find shallower density
profiles, in some cases with nearly constant-density central cores~\cite[e.g.,][]{Oh2011,Oh2015,Newman2013,Adams2014}.
This cusp-core problem has persisted for 25 years despite major improvements
in both simulations and observations.  Currently-favored explanations
for the problem include modifications to the gravitational potential
of galaxies by repeated episodes of strong stellar feedback~\cite[e.g.,][]{Governato2012} or self-interacting dark matter models~\citep[e.g.,][]{SS2000,Kaplinghat2016}.  Simulations also
predict that the dark matter halo mass function rises steeply
($dN/dM \propto M^{-1.9}$; \citealt{springel2008}) with decreasing mass
down to the free-streaming scale ($\sim10^{-6}$~M$_{\odot}$ for GeV-mass dark matter particles).  On the other hand, the
galaxy luminosity function has a significantly shallower slope, and if
all low-mass dark matter halos above $\sim10^{8}$~\msun\ contain galaxies
then the Milky Way hosts many fewer satellite galaxies than
expected~\citep{Klypin1999,Moore1999}.  While this missing
satellite problem can be explained by restricting galaxy formation to
specific subsets of the dark matter halo population (e.g., \citealt{kravtsov2010}), the existence of halos with masses smaller than those of
galaxies is a fundamental --- and testable --- prediction of CDM
(e.g., \citealt{bullock2017}).  Thus, although there are
plausible explanations for these problems without invoking a crisis
for the $\Lambda$CDM model, new observations are required to verify
the proposed solutions and demonstrate that the predictions of
$\Lambda$CDM on small scales are accurate.

\vspace{-0.4cm}
\subsection{Testing $\Lambda$CDM}
\vspace{-0.1cm}

\textbf{Local volume:} Measurements of dark matter density
profiles in the nearby universe have generally focused on low-mass
disk galaxies, which contain significant baryonic components.
Episodes of star formation in
these systems may substantially alter the original distribution of
dark matter \citep[e.g.,][]{DiCintio2014}, making the measurements
difficult to interpret in the context of the prediction of central
cusps.  Ideally, one would prefer to determine the density profiles of
the most dark matter-dominated objects known, namely the dwarf
galaxies orbiting the Milky Way.  The smallest of these galaxies are
so dark matter-dominated that stellar feedback is unlikely to have
affected their dark matter halos on $\sim 100$~pc scales (e.g., \citealt{bullock2017}).
Many studies have attempted to constrain the dark matter distribution
in Milky Way dwarf spheroidals via line-of-sight velocity
measurements, but the degeneracy between the mass profile and the
orbital anisotropy of the stars has led to conflicting results \citep[e.g.,][]{Battaglia2008,Amorisco2012,Breddels2013,Strigari:2014yea}.

Similarly, a number of methods have been used to attempt to detect the
dark (or nearly dark) subhalos predicted by \lcdm\ in the Local Group
and surrounding regions.  These techniques include searches for
galaxies associated with low-mass gas clouds \cite[e.g.,][]{SB2002,Adams2016}, searches for gaps in tidal streams~\cite[e.g.,][]{Carlberg2012,Bonaca2018}, and analyses of their visible
gravitational signatures on the outer gas disks of
galaxies~\citep{CB2009,Chakrabarti2011}.  While
each of these approaches continues to hold promise, it is not clear at
present whether any of them will yield definitive results.

\textbf{Cosmological distances:} Strong gravitational lensing offers a
unique way to detect low-mass dark matter halos at cosmological
distances~\citep[e.g.,][]{MS1998,MM2001,DK2002,Koopmans2005}. Because gravitational lensing is affected only by the projected
mass surface density, it is sensitive to mass concentrations
independent of their baryonic content. Thus, purely dark halos can be
detected, providing a stringent test of dark matter
models.  Identified structures include both subhalos
within the primary lensing galaxy (``substructure'') and halos along
the line of sight between the lensed object and the observer (``LOS
structure''; \citealt{Li2017,Despali2018}).

Lensing substructure can alter both the distortion of extended lensing arcs
and the magnification of unresolved sources.  Induced distortions in
extended Einstein rings, a method known as gravitational imaging, have
resulted in the detection of substructure with inferred lensing masses
of $M_{\rm lens} \approx 10^{8-9}$~M$_{\odot}$ with HST, Keck AO and
ALMA data \citep{Vegetti_2010_2,Vegetti_2012,2014MNRAS.442.2017V,Hezaveh_2016_2,Birrer2017,Ritondale:2018}.  The sensitivity limit of
a lensing detection is set by the angular resolution of the data
and the structure of the lensed source. Additional modeling challenges arise as a result of the complexity of the data
set, uncertainties in telescope and instrument response and
degeneracies in the lens model and source structure.

Unresolved magnification effects have been probed by examining the
flux ratios between neighboring images of multiply-imaged
quasars. These data are consistent with CDM at scales of $\ga
10^{9}$~M$_{\odot}$ \citep{DK2002,Nierenberg:2017}.
However, existing observations lack sufficient resolution to probe interesting regimes of viable warm dark matter models, which require
reaching $\sim 10^7$~M$_\odot$.
The magnification ratio and thus the interpretation of the flux ratios
depends on the size of the emitting source region (radio vs.~narrow-line emission region) and on the smooth component of the lens
model with respect to which anomalies can be
identified. Over-simplified lens models can bias the expected flux
ratios \citep{Xu:2015dra}, and baryonic structures can affect the
interpretation of anomalous structures in both real lenses \citep{Hsueh:2016aih,Hsueh:2017zfs,Hsueh:2017nlk} and simulated ones \citep{Gilman:2016uit}.
Exploiting the full power of lensing to constrain the nature of dark
matter will require both sophisticated
analysis tools {\bf and} exquisite data, which can only be obtained with the angular resolution
and collecting area of extremely large telescopes (ELTs).

\vspace{-0.5cm}
\section{An Observational Path Forward}
\vspace{-0.3cm}


\textbf{Dwarf galaxy kinematics:} Combined radial velocity and proper motion measurements for  stars in a dwarf galaxy would tightly constrain the
stellar orbits within the galaxy and break the degeneracies that have plagued previous
density profile studies.  Obtaining such 3D motions requires both
spectroscopy and high-precision astrometry.  Theoretical modeling
suggests that velocity measurements accurate to $\sim3$~\kms\ in each dimension for a sample of $\sim300$ stars is the minimum
necessary to reliably recover the gravitational potential~\citep[e.g.,][]{Strigari2007}.  Proper motions of faint
dwarf galaxy stars at the required accuracy can only be measured with
laser guide star AO imaging.  Here the AO field of view is
critical, because the surface density of member stars at the relevant
magnitudes is $<0.1$~arcsec$^{-2}$, requiring large fields to obtain the sample
needed to determine the tangential velocity dispersion.
A velocity uncertainty of 3~\kms\ translates to a proper motion
uncertainty of $6.3~(100~{\rm kpc}/d)$~$\mu$as~yr$^{-1}$, or
7.9~(21.0)~$\mu$as~yr$^{-1}$ at a distance of 80~(30)~kpc.  With the
anticipated $\sim15$~$\mu$as astrometric error floor of a 30~m
telescope \citep{Wright2016SPIE}, these proper motions could be measured
over a time baseline of a few years.

Radial velocities for dwarf galaxy member stars can be obtained with
multi-object spectrographs on large ($>6$~m) telescopes.  
Velocity measurements for faint stars have been demonstrated at the
1.5~\kms\ level at $R = 6000$~\citep[Keck/DEIMOS;][]{Kirby2015} and the
1.0~\kms\ level at $R = 12000$~\citep[Magellan/IMACS;][]{Simon2017}
with existing instruments.  We recommend that future spectrographs on
large telescopes (1) plan to incorporate gratings that will
provide a spectral resolution of at least $R=6000$ at the wavelength
of the Ca triplet absorption lines ($\sim8500$~\AA) and/or the Mg~b
triplet ($\sim5200$~\AA), and (2) are designed to maximize
stability.  Milky Way satellite galaxies typically have half-light
radii of $\sim10\arcmin$, so the larger the field of view and
multiplexing that can be achieved, the more efficiently the
observations can be obtained.

\begin{SCfigure}[][h]
\includegraphics[width=0.45\textwidth]{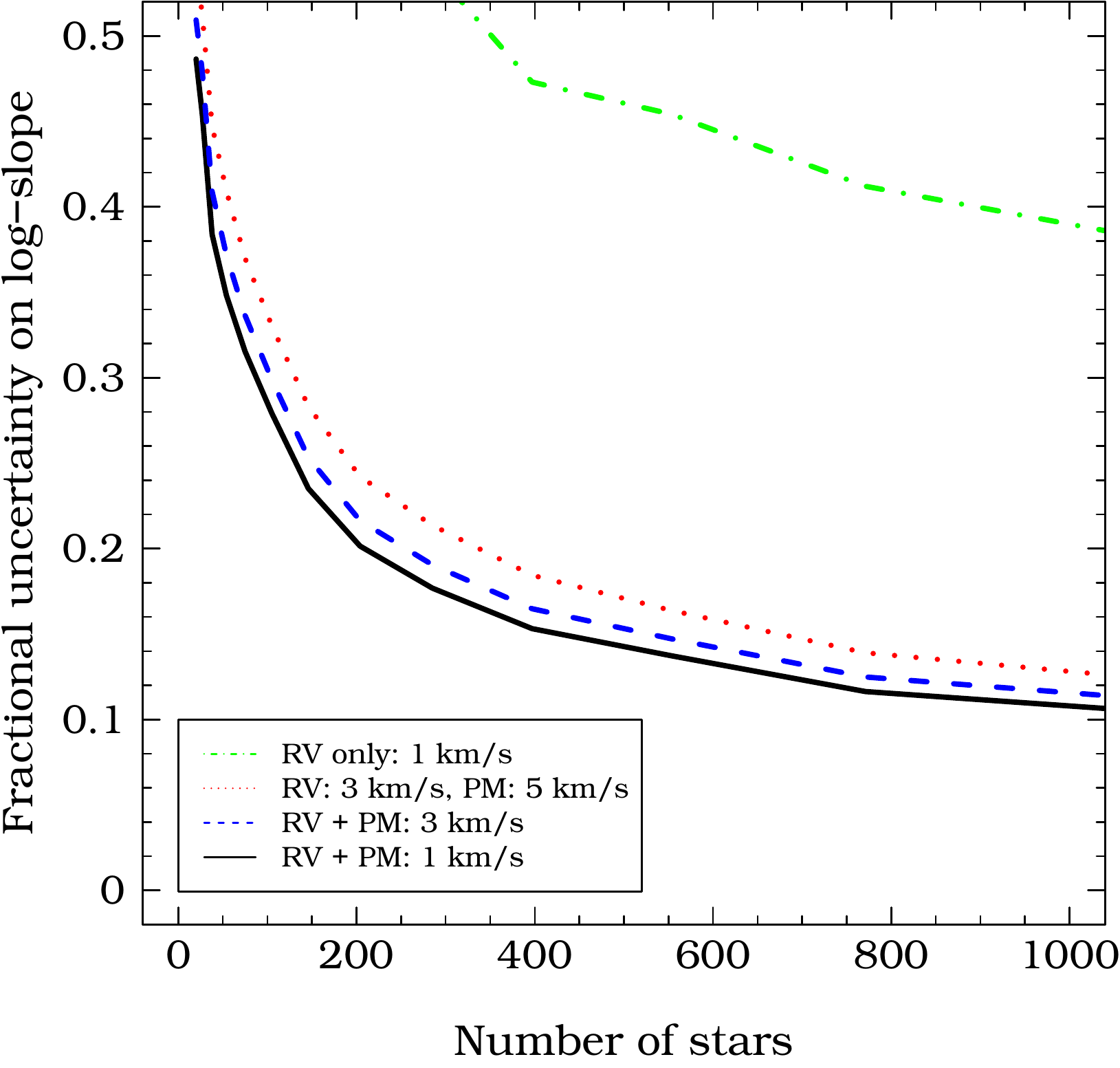}
\caption{Simulated recovery of the dark matter density profile of a
  dwarf spheroidal via stellar kinematic data (based on the results of~\citealt{Strigari2007}).  Including 3D stellar velocities (red/blue/black curves) dramatically reduces the
  uncertainty on the central slope of the density profile relative to
  a data set consisting only of radial velocities (green curve).  For a sample of
  300 stars with proper motion and radial velocity uncertainties of
  3~\kms, the expected measurement uncertainty on the slope is 0.2,
  enabling a 5$\sigma$ detection of a central density cusp. \vspace{0.20in} }
\label{fig:dwarf}
\end{SCfigure}

A key ELT goal should be to measure the radial
velocities and proper motions of 300 stars per galaxy in several
Milky Way satellites, with a typical accuracy per star of
3~\kms\ (Fig.~\ref{fig:dwarf}).
These measurements will directly determine the velocity anisotropy of
the stellar orbits within each dwarf, enabling tight constraints to be
placed on the inner density profiles of their dark matter halos.
Observations of multiple dwarf galaxies will determine the range of
halo profiles that exist and avoid the possibility of being misled by
the unique history of any individual galaxy.

\vspace{-0.1in}
The ideal targets for this experiment should satisfy the following
criteria: (1) Low stellar mass ($<10^{6}$~\msun), to minimize stellar feedback effects on the galaxy's mass distribution; (2) High stellar surface density, to minimize the number
of pointings needed to reach a sample of 300 proper motion
measurements; (3) Large velocity dispersion, to increase the expected
proper motion signal; (4) Small distance, to increase the expected
proper motion signal and maximize the brightness of each star; and (5)
An orbit that does not approach the Milky Way too closely, to minimize
the impact of Galactic tides on the structure of the dwarf.  
Clear choices include Draco and Ursa Minor in the north, and Sculptor and Carina in the south.
Simple ELT exposure time estimates indicate that the required tangential
velocity accuracy of 3~\kms\ can be obtained with $\sim16$~hr
integrations per pointing and a time baseline of 5 years.  Imaging of
5 pointings per galaxy would provide a sample size of 300 stars in
each target.  The dark matter density profile of one dwarf could
therefore be measured with a total investment of $\sim160$~hrs.

ELTs offer several key advantages over space-based observing platforms
for this project.  In particular, the angular resolution of ELTs is an
order of magnitude better than that of HST at the same wavelength.
The demonstrated astrometric performance of
HST relative to its diffraction limit is comparable to that expected
for ELT imagers, providing a factor of $\sim5$ advantage for ELTs.
Furthermore, the enormous increase in collecting area means that ELTs
can provide significantly larger stellar samples in a dwarf galaxy by
detecting fainter stars at high S/N.  On the other hand, the HST field
of view is a factor of $\sim35$ larger than planned first-generation
ELT instruments, providing a significant increase in efficiency in the
case where a large area must be observed.  For the dwarf spheroidal observations
described above, the proper motion measurement that can be obtained in
5~years with an ELT would require a time baseline of $\sim25$~years
with HST.

\textbf{Strong gravitational lensing:} Constraining dark matter using strong lens systems requires
high-sensitivity and high-angular resolution imaging and
spectroscopy. For the gravitational imaging technique, comparisons of
HST vs.~Keck AO imaging data have confirmed that sensitivity scales
with image resolution.
The current AO system on Keck produces typical angular resolutions of
$60-90$~mas and a Strehl ratio of $\sim10-30$\% for off-axis
targets. With a bigger mirror, a better AO system, and a larger
population of potential tip-tilt stars, ELTs will perform
significantly better. The AO systems on next-generation telescopes are
designed to reach Strehl ratios as high as 90\% in the K
band. Furthermore, they will include built-in PSF reconstruction
software that provides a model of the PSF for data analysis. The
combination of improvements in resolution, sensitivity, and PSF
quality and knowledge will make the ELTs much more capable than
current systems. Specifically, only the brightest handful of lenses in
the sky can currently be targeted with $8-10$~m class telescopes.  The
US ELT program, on the other hand, will provide virtually full sky
access, allowing us to target large samples of rare quadruply-imaged
objects, selecting the ones with the best (i.e., the ones containing
the most structure) extended images for the gravitational imaging
technique.  Moreover, Fisher analyses of simulated data have shown
that lens modeling of spectroscopically resolved images can mitigate
the source-subhalo degeneracies, drastically increasing the
sensitivity of the observations to low-mass subhalos \citep{Hezaveh2013}.

For the flux-ratio anomaly technique, a very promising route forward
is AO-assisted IFU spectroscopy.  Integral field spectroscopy allows
the measurement of flux ratios from the narrow-line region of the AGN,
which should be immune to microlensing \citep[see][]{Nierenberg:2014aa,Nierenberg:2017}. Unfortunately the technique is currently restricted to only a
handful of quads, given the limitations of tip-tilt stars and the need to
have narrow lines fall in transparent windows of the Earth's atmosphere.
As in the case of gravitational imaging, a two-hemisphere US ELT system will
enable the application of the technique to the samples required to
reach definitive conclusions on the nature of dark matter by virtue of
their sensitivity (exploiting the $D^4$ advantage of point sources in
diffraction-limited mode), resolution (for astrometry of images), and
full-sky access.

In order to measure the dark matter halo mass function in the range
$10^{7} - 10^{10}$~M$_{\odot}$, forward modeling simulations indicate
that samples of $\sim50$ extended arcs and $\gtrsim50$
quadruply-imaged quasars would be needed.  The former provide
sufficient statistics of the rarer $10^{9-10}$~M$_{\odot}$
substructures, while the latter are needed for direct and statistical
constraints on the mass function in the range $10^{7-9}$~M$_{\odot}$.
For the gravitational imaging technique, the brighter the source
galaxy, the more clumpy its structure, and the better the separation
between the lens galaxy and the background lensed object, the better
the substructure constraints will be.  The sample of lenses should
contain low and high redshift lenses (and sources) to statistically
break the degeneracy between LOS structure and subhalos bound to the
main deflector and characterize possible evolutionary trends.
Gravitational imaging will be able to provide additional constraints
on the inner density slope of the massive subhalos
($\sim10^{10}$~M$_{\odot}$).  Observations of each lens system are
expected to require integration times of a few hours on an ELT.

\begin{figure}[t!]
\includegraphics[width=1.0\textwidth]{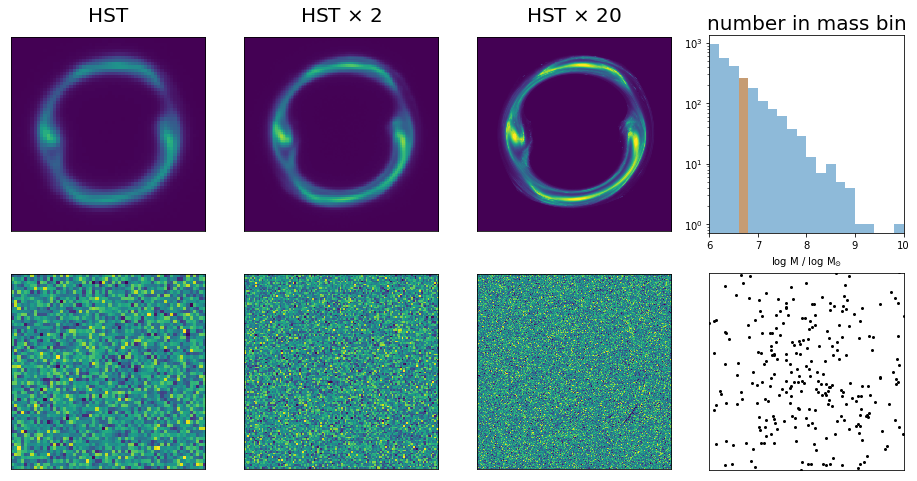}
\vspace{-0.4in}
\caption{The effect of lensing substructure on a resolved arc for
  three different imaging resolutions. The baseline (left panel)
  illustrates an HST image with FWHM 0.1\arcsec, followed by a ground
  based AO observation of FWHM 0.05\arcsec\ (second panel) and an
  ELT-like observation with FWHM 0.005\arcsec\ (third panel). Top row:
  simulated images. Bottom row: residuals per pixel (in S/N) between a
  simulation with the substructure signal and without the presence of
  lensing substructure. For this particular simulation, we only
  included substructure in the mass range $10^{6.6-6.8}$~M$_{\odot}$.
  The imprint of these low-mass structures is only detectable in the
  ELT simulation (third panel, bottom row).\vspace{-0.15in} 
\label{fig:lensing}
}
\end{figure}

\vspace{-0.6cm}
\section{Summary}
\vspace{-0.4cm}

We suggest that two complementary ELT experiments could provide
fundamental tests of whether dark matter is cold and non-interacting by means of (1) the 3D motions of stars in nearby dwarf galaxies and (2)
gravitational lensing of distant galaxies and quasars.  The dwarf
galaxy component of this program would employ diffraction-limited
imaging (combined with deep spectroscopy) to measure the 3D velocities
of $\sim300$ stars in each of two Milky Way satellite galaxies.  The lensing component of the program would focus on diffraction-limited, high S/N IFU spectroscopy of 50 multiply-imaged quasars and 50 resolved arcs.  Investment of $\sim600$~hrs of ELT time in these projects would result in unparalleled constraints on the dark matter density profiles of undisturbed halos and the subhalo mass function on scales well below those probed by luminous structures.

\pagebreak

\bibliographystyle{apj}

\end{document}